\input mtexsis
\hoffset=-0.35cm
\voffset=1.3cm
\headline={{\underbar{\it Metaphysical Review,
Vol. 3, no. 1, July 1996}}\hfil {\underbar{\it www.meta.unh.edu}}}
\footline={\hfil\folio\hfil}
\center
\Tbf{The Plight of `I Am'}
\endcenter

\center
Joy Christian
\endcenter

\center
Wolfson College, University of Oxford, Oxford OX2 6UD, United Kingdom
\endcenter

\bigskip
\bigskip

\parindent 0pt
\parskip 0.40cm
\baselineskip 0.55cm

Once upon a time ........\n
.......................................there was a French soldier called
Descartes.

One fine evening, as he was passing through a battle ground, he saw `I am'
sitting happily on the wall of sound metaphysics. 

He gazed at it for a while, and --- after a bit of thinking --- proclaimed:

\center
``{\underbar{I {\it think}, therefore I am ({\it i.e.}, I exist)}}'' (Discourse on Method, 1637).
\endcenter

`I am' smiled at him, easing itself into its comfortable classical
seat on the tall wall.

Three hundred years went by. Empires declined and fell,
and empires were born.\n
Yet, nothing really deterred `I am' from its privileged pedestal.\n
Nothing, that is, that was as momentous as what was to happen.

Then, in 1900, Planck glimpsed the Quantum.
Once refined by Heisenberg,
Schr\"odinger, and Dirac, this Quantum lead to the great fall of `I am'.
For a quantal `I am' is merely a `potential' (or indefinite) `I am' and not
necessarily an `actual' (or definite) `I am'.

Off it went tumbling down the wall of sound metaphysics, utterly bemused.

As luck would have it, a brave and mighty
knight called Bohr was passing by, just in time to save it --- so he thought.
While still riding on his unflinching classical horse, he charged in with his
cutting quantal sword and decreed:

\center
``{\underbar{I am {\it classical}, therefore I am}}'' (Copenhagen, 1935).
\endcenter

But, clearly, his decree did not have the right ring to it. For the classical
could not be distilled from the quantal. This, in fact, was the {\it very
reason} for the plight of `I am'! 

Yet, Bohr found many followers in his crusade to save `I am' ---
all happy and content, if not complacent. And his decree might have been final,
had it not been for the heretics like Einstein, Bohm, Wigner, and Everett.

With some help from his ingenious confederate von Neumann, Wigner sought to
mend the weakness of Bohr's decree. He reached out deep into his own psyche
and surmised:

\center
``{\underbar{I {\it am (conscious)}, therefore I am}}'' (Princeton, 1961).
\endcenter

But this
did not sound tautologous perhaps only to the God of Moses (Exodus 3:14).

For Bohm, on the other hand, the quantal was too much. 
He was quite happy to be classical, even if he would have to remain `hidden'
for it. So, in the face of Bohr's decree, he dared divulge his scheme:

\center
``{\underbar{I {\it cloak (and uncloak only non-locally)}, therefore I am}}'' (London, 1952).
\endcenter

Although Lord Krishna got away with such a specious trick (Bhagavad-Gita
7:24-25, 8:18-21, 9:4-5), Everett was clearly opposed to it ---
vehemently opposed to it. For he preferred to be purely quantal all the way!
No, no, no, David, he exclaimed,

\center
``{\underbar{I {\it split}, therefore I am}}'' (Princeton, 1957).
\endcenter

(Apologies to Karel Kucha\v r; cf. {\it The Garden of Forking Paths}, Jorge
Luis Borges, 1941.)

This splitting did a lot of good to Captain Kirk as he
boldly took his star-ship Enterprise where no man had gone before.

But others did not feel like splitting --- or cloaking, for that matter.

And then there was this curious camp in the battle field, still in awe of Bohr.
One of the many puissant
emissaries of this largest and oldest camp was Gell-Mann.
Reminiscent of Bohr's decree, he appealed to the environment and maintained:

\center
``{\underbar{I {\it decohere}, therefore, FAPP, I am}}'' (DLP, 1962 - Omn{\`e}s, 1994).
\endcenter

\rightline{(Apologies, again, to Karel Kucha\v r.)}

This seemed to do a lot of good for quantum cosmology, if not for quantum
gravity.

Well, perhaps, frowned the mutineers, who found no prudence in decoherence.
They preferred the genuine `I am' and not a {\sl FAPP} `I am'. A FAPP `I am',
they cried out, is still only a `potential' `I am', not an `actual' `I am' ---
a FAPP `I am' is no `I am'!

And, so, they longed for more than just decoherence --- so much more that they
established a
small outpost of their own, and conspired to {\it fudge} the Quantum.
The commander-in-chief of this small but worthy faction was Ghirardi.
From the sanctuary of their godfather John Bell, Ghirardi
gathered his troops together and declared:

\center
``{\underbar{I {\it spontaneously localize}, therefore I am}}'' (Italy, 1986).
\endcenter

But their fudge remained {\it ad hoc} as their physics remained obscure.
And most embarrassingly, there also remained the `tails' of Schr\"odinger's
Cat. They just would not go away. Einstein surely would not have liked this
--- neither would have Descartes, for that matter.

The tails did not bother Bell though. When `I am' questioned him about them at
a talk at MIT in 1990 shortly before his premature death, he quipped in his
characteristic Irish tone: ``Your worries are irrational.''

They did bother some, however, like Albert and Loewer (1990, 1996), who would
rather have their minds split {\it \`a la} Everett than have these tails dawdle
around. And even Shimony --- a
staunch partisan of the fudge --- was somewhat concerned about `I am' if,
despite a cleverer fudge, the tails were to remain (1991).

Others murmured that {\it Gravity}
was the culprit fudging the Quantum. The most
prominent voice among these was that of Penrose. Unfortunately, he appeared to
be more concerned about ``orchestrating'' the state of the elusive conscious
rather than the state of the poor `I am'. Nevertheless, as one of the central
strategies for accomplishing his primary goal to fathom the conscious,
he contended:

\center
``{\underbar{I {\it quantum-gravitate}, therefore I am}}'' (Oxford, 1989, 1994).
\endcenter

He would rather quantum-gravitate and, as a result, {\it non-algorithmically}
spontaneously localize than just boringly decohere.

But, again, no one was there to eradicate the tails. Not even
Gravity the Exotic.
Surely, the fudge was so constructed that the tails were able to fool Physics;
and, thus, it was possible to relocate `I am' on the wall of almost sound
physics. But, of course, that is not where the poor `I am' belonged;
and Metaphysics was not going to be fooled by such an obtuse trick.
For, to Metaphysics, the tails were as monstrous as the Cat itself!

Alas! It was not possible to put `I am' back on the wall of sound metaphysics
even with the help of Gravity. Of course, Gell-Mann alone --- or DLP (1962) or
Hepp (1972), for that matter, long before him --- could have put `I am' on the
wall of almost sound
physics --- and without resorting to the fudge! But that was not the place for
`I am'.

If `I am' was to {\it remain} `I am', it {\it had} to be reestablished on the
wall of sound metaphysics, not just physics --- albeit with the help of fair
Physics. So, ...............

......was `I am' an offshoot of a cleverer version of
{\it spontaneous-localization}
due to {\it quantum-gravitation}? Or was it an intricate manifestation of a
yet-to-be-discovered much more elegant and subtle {\it non-computable} facet of
the unknown quantum theory of gravity?

`I am' knew not. 

For, as long as the tails of the Schr\"odinger's Cat lingered, `I am' was no
`I am'. 

Thus, for now, `I am' lay shattered at the bottom of the wall it
sought to top, as all the Queen's horses and all the Queen's men toil to
put the poor `I am' back together again. ${\diamondsuit}$

\vfill\eject

{\Tbf References}

\item{1.}Ren\'e ~Descartes, {\it Discourse on the method of rightly conducting
one's reason and reaching the truth in the sciences}, (1637).

\item{2.}Niels ~Bohr, {\it Physical Review}, vol. 48, p. 696 (1935).

\item{3.}Eugene ~Wigner, in {\it The Scientist Speculates}, edited by I.J.
Good, p. 284 (Heinemann, London, 1961).

\item{4.}David ~Bohm, {\it Physical Review}, vol. 85, p. 166 (1952).

\item{5.}Hugh ~Everett III, {\it Reviews of Modern Physics}, vol. 29, p. 454
(1957).

\item{6.}A. ~Daneri, A. ~Loinger, and G. ~Prosperi (DLP), {\it Nuclear
Physics}, vol. 33, p. 297 (1962).

\item{7.}Ronald ~Omn\`es, {\it The interpretation of quantum mechanics}
(Princeton University Press, Princeton, New Jersey, 1994).

\item{8.}G. ~Ghirardi, A. ~Rimini, and T. ~Weber, {\it Physical Review}, vol.
D 34, p. 470 (1986).

\item{9.}David ~Albert, in {\it Sixty-Two Years of Uncertainty}, edited by
A. ~Miller, p. 153 (Plenum Press, New York, 1990).

\item{10.}D. ~Albert and B. ~Loewer, in {\it Perspectives on Quantum Reality},
edited by R. ~Clifton, p. 81 (Kluwer Academic Publishers, The Netherlands,
1996).

\item{11.}Abner ~Shimony, {\it Search for a naturalistic world view}, vol. II
(Cambridge University Press, Cambridge, 1991) p. 55.

\item{12.}Roger ~Penrose, {\it Emperor's new mind} (Oxford University Press,
Oxford, 1989).

\item{13.}Roger ~Penrose, {\it Shadows of the mind} (Oxford University Press,
Oxford, 1994).

\item{14.}K. ~Hepp, {\it Helv. Phys. Acta.}, vol. 45, p. 237 (1972).

\end